\documentstyle[twoside,fleqn,espcrc2]{article}

\title{
\hfill
\parbox{3cm}{\normalsize DPNU-99-29 \\
{\tt  hep-lat/9909065}
}\\
\vspace{0.1cm}
Low energy effective action of domain-wall fermion \\
and the Ginsparg-Wilson relation
}

\author{Y. Kikukawa\address{Department of Physics, Nagoya University, 
Nagoya 606-8602, Japan}%
        \thanks{
This talk is based on the collaborations with 
T.~Noguchi and T.~Aoyama (Kyoto Univ.).
These works are supported in part by Grant-in-Aid for Scientific 
Research from Ministry of Education, Science and 
Culture of Japan. (\#10740116,\#10140214)
}
}
\begin{document}

\begin{abstract}
We derive the effective action of the light fermion field 
of the domain-wall fermion, which is referred 
as $q(x)$ and $\bar q(x)$ by Furman and Shamir. 
The inverse of the effective Dirac operator 
turns out to be identical 
to the inverse of the truncated overlap Dirac operator except a local 
contact term, which would give the chiral symmetry breaking 
in the Ginsparg-Wilson relation. 
We argue that there are direct relations between
the low energy observables of the domain-wall QCD
and observables of the Ginsparg-Wilson fermion 
described by the (truncated) overlap Dirac operator.

\end{abstract}
 
\maketitle

\section{Introduction}

Recently, our understanding of chiral symmetry on the lattice
has substantially improved. 
Lattice Dirac operators 
have been obtained \cite{overlap-D,fixed-point-D},
\begin{equation}
  S_F = a^4 \sum_x \bar \psi(x) D \psi(x) ,
\end{equation}
which are gauge covariant, 
define local actions \cite{locality-of-overlap-D} and 
satisfy the Ginsparg-Wilson relation \cite{ginsparg-wilson-rel},
\begin{equation}
  \gamma_5 D + D \gamma_5 = a D R \gamma_5 D .
\end{equation}
The Ginsparg-Wilson relation 
implies the exact chiral symmetry of the action
under the transformation \cite{exact-chiral-symmetry}
\begin{equation}
\label{eq:chiral-transformation-of-psi}
  \delta \psi(x) = \gamma_5 \left( 1-aRD \right) \psi(x), \quad
  \delta \bar \psi(x) = \bar \psi(x) \gamma_5 . \nonumber\\
\end{equation}

The goal of this paper is to understand the chiral property
of the light fermion of the domain-wall QCD
from the point of view of the exact chiral 
symmetry based on the Ginsparg-Wilson 
relation.
The domain-wall fermion \cite{domain-wall-fermion} is the basis 
of Neuberger's overlap Dirac operator \cite{overlap-D}.
In a simplified formulation 
\cite{boundary-fermion,boundary-fermion-QCD},
the domain-wall fermion consists of $N$-flavor Wilson 
fermions with a certain flavor-mixing mass matrix.
Due to its structure of the chiral hopping and the boundary condition 
in the flavor space, a single light Dirac fermion can emerge in the 
spectrum. This light fermion can be probed suitably
by the field variables at the boundary in the flavor space, 
which are referred as $q(x)$ and $\bar q(x)$ by Furman and Shamir:
\begin{eqnarray}
  q(x) &=& \psi_{1L}(x) + \psi_{NR}(x) , \nonumber\\ 
 \bar q(x) &=& \bar \psi_{1L}(x) + \bar \psi_{NR}(x).
\end{eqnarray}
It has been argued that the chiral symmetry of 
the light fermion is preserved up to corrections suppressed 
exponentially in the number of flavors 
\cite{boundary-fermion-QCD}.
The chiral transformation of the light fermion field
in this context is defined as
\begin{equation}
\label{eq:chiral-transformation-of-q}
\delta q(x) = \gamma_5 q(x), \quad
\delta \bar q(x) = \bar q(x) \gamma_5 .
\end{equation}

It has been known that 
in the subtraction scheme proposed by Vranas \cite{vranas-pauli-villars},
the partition function of the domain-wall fermion 
reduces to a single determinant of the truncated overlap 
Dirac operator \cite{truncated-overlap}:
\begin{equation}
\label{eq:truncated-overlap-dirac-operator}
D_N= 
\frac{1}{2a}\left(
1+\gamma_5 \tanh \frac{N}{2} a_5 \widetilde H \right),
\end{equation}
where $\widetilde H$ is defined through the transfer matrix of the
five-dimensional Wilson fermion with a negative 
mass. 
(In this derivation,
the positivity of 
\begin{equation}
B = 1 + a_5 \left( 
-\frac{a}{2} \nabla_\mu\nabla_\mu^\ast - \frac{m_0}{a} \right)
\end{equation}
is required for the transfer matrix to be defined consistently.
It is assured when $ 0 < \frac{a_5}{a} m_0  < 1$ . It is also assumed that
$N$ is even. )

In this paper, we will show that 
there is a simple correspondence 
between the light fermion field of the domain-wall fermion and 
the Dirac field described by the (truncated) overlap Dirac 
operator:
\begin{eqnarray}
q(x), \, \bar q(x) 
&\Longleftrightarrow& 
\left(1-\frac{a}{2}RD_N \right) \psi(x) ,  \, \bar \psi(x) .
\end{eqnarray}
We will then discuss the relation between
the low energy observables of the domain-wall fermion 
and those of the Ginsparg-Wilson fermion 
described by the (truncated) overlap Dirac 
operator. \cite{kikukawa-noguchi}

\section{Effective action of the light fermion}

For this purpose, we first derive the low energy effective action 
of the light fermion field by integrating out 
$N-1$ heavy flavors of the domain-wall fermion:
\begin{equation}
  S_N^{\rm eff} = a^4 \sum_x \bar q(x) \, D_N^{\rm eff} \, q(x). 
\end{equation}
Since the inverse of this effective Dirac operator should give
the propagator of $q(x)$ and $\bar q(x)$, 
this can be achieved by calculating the propagator.
The result turns out that \cite{kikukawa-noguchi}
\begin{eqnarray}
D_N^{\rm eff}(x,y) 
&=& 
\frac{1}{a^4}\left\langle q(x) \, \bar q(y) \right\rangle ^{-1} \nonumber\\
&=& \frac{1}{a_5} \, 
\frac{1+\gamma_5 \tanh  \frac{N}{2} a_5 \widetilde H}
     {1-\gamma_5 \tanh \frac{N}{2} a_5  \widetilde H} .
\end{eqnarray}

This result implies that 
the inverse of the effective Dirac operator 
gives the inverse of the truncated overlap Dirac operator up to 
a local contact term:
\begin{eqnarray}
\label{eq:effective-D-vs-truncated-D}
\frac{a}{a_5} { D_N^{\rm eff} }^{-1}+ a \delta(x,y) 
&=& 
{ D_N^{\rm \phantom{f}} }^{-1} .
\end{eqnarray}
This contact term just takes account of the 
chiral symmetry breaking in the Ginsparg-Wilson relation, 
which holds true for the overlap Dirac operator 
in the limit of the infinite 
flavors.

\section{Relation between $q(x)$ $\bar q(x)$ and $\psi(x)$ $\bar \psi(x)$}
After integrating out $N-1$ heavy flavors of the domain-wall fermion 
and also corresponding $N-1$ flavors of the Pauli-Villars boson, 
the partition function of the domain-wall fermion
assumes the following expression in functional integral:
\begin{eqnarray}
\bar Z_{\rm DW}
&=& \int [d q d \bar q] [d Q d \bar Q] e^{- \bar S_N^{\rm eff}[q,\bar
  q, Q, \bar Q] } ,
\end{eqnarray}
where we have denoted 
the field variables of the Pauli-Villars fields 
at the boundary in the flavor space by $Q(x)$ and $\bar Q(x)$. 
The total effective action is given 
by
\begin{eqnarray}
\label{eq:effective-action-subtracted}
\bar S_N^{\rm eff}[q,Q]
&=& a^4 \sum_x \bar q(x) \, D_N^{\rm eff} \, q(x)  \\
&+& a^4 \sum_x \bar Q(x) \left\{ D_N^{\rm eff}+\frac{1}{a_5} \right\}
Q(x). \nonumber
\end{eqnarray}
Note that $Q(x)$ and $\bar Q(x)$ 
acquires the unit mass of order $1/a_5$ 
because of the anti-periodic boundary condition in the flavor space.
Since Eq.~(\ref{eq:effective-D-vs-truncated-D}) implies 
the following identity
\begin{equation}
\label{eq:truncated-overlap-vs-effective-action-q}
\frac{ a_5 D_N^{\rm eff} }{1+ a_5 D_N^{\rm eff}}
= a D_N ,
\end{equation}
we see that the partition function can be rewritten into
\begin{eqnarray}
\bar Z_{\rm DW}
&=& \int [d \psi d \bar \psi] \, 
             e^{- a^4 \sum_x \bar \psi(x) D_N \psi(x) } 
\end{eqnarray}
through the change of the field variables along the relation
\begin{eqnarray}
\label{eq:relation-q-psi}
q(x)&=& Z \, \frac{1}{1+a_5 D_N^{\rm eff}} \, \psi(x) \nonumber\\
&=&Z \, \left( 1-\frac{a}{2} R D_N \right) \psi(x) ,  \nonumber\\
\bar q(x)&=& \bar \psi(x), \qquad 
Z=\frac{a_5}{a}, \, R=2.
\end{eqnarray}
The Jacobian associated with the change of the field variables
just compensates the determinant resulting from the integration 
of $Q(x)$ $\bar Q(x)$.
Thus we reproduce the result of \cite{truncated-overlap}.

An immediate consequence of Eq.~(\ref{eq:relation-q-psi}) 
is that any local observables of the domain-wall fermion 
written in terms of only $q(x)$ and $\bar q(x)$ 
can be related 
to local observables written in terms of $\psi(x)$, $\bar \psi(x)$ and
$D_N$, as long as $D_N$ assumes a local Dirac operator:
\begin{equation}
  {\cal O}_{\rm DW}^{(N)}[q,\bar q]
 ={\cal O}_{\rm GW}^{(N)}[\psi, \bar \psi; D_N ] .
\end{equation}
Then we can relate the chiral properties of these observables
to the (would-be) exact chiral symmetry based 
on the Ginsparg-Wilson relation.
A typical example is given by the chiral multiplet of scalar and
pseudo scalar bilinear operators:
from Eq.~(\ref{eq:relation-q-psi}),
we obtain a relation among them,
\begin{eqnarray}
\bar q(x) q(x) 
&=& Z \, \bar \psi(x) \left(1- \frac{a}{2}RD_N \right)\psi(x) , 
\nonumber\\
\bar q(x) \gamma_5 q(x) 
&=&Z \, \bar \psi(x) \gamma_5 \left(1- \frac{a}{2}RD_N \right)\psi(x).
\nonumber
\end{eqnarray}
Note that 
operators in the r.h.s.
consist the exact chiral multiplet of the chiral transformation 
Eq.~(\ref{eq:chiral-transformation-of-psi}) 
in the limit of infinite flavors $N \rightarrow \infty$, 
as discussed by Niedermayer \cite{neidermayer-lat98}.

\section{Axial-vector current and chiral anomaly}

Let $\bar A^a_{\mu {\rm DW}}(x)$ and 
$ \widetilde A^a_{\mu {\rm GW}}(x)$ denote the axial-vector currents of 
the domain-wall fermion and of the Ginsparg-Wilson fermion 
described by Neuberger's Dirac operator with $\widetilde H$, 
respectively. 
An important aspect of $\bar A^a_{\mu {\rm DW}}(x)$ we should note 
is that it is written not only in terms of the light fermion field, but
also in terms of all heavy flavors including the Pauli-Villars fields.
Using Eq.~(\ref{eq:effective-D-vs-truncated-D}), 
it is possible to derive the following identities which relate 
them:
\begin{eqnarray}
\label{relation-of-axial-vector-currents-q}
&& \lim_{N\rightarrow \infty}
\left\langle \bar A^a_{\mu {\rm DW}}(x) \right\rangle 
=\left\langle  \widetilde  A^a_{\mu {\rm GW}}(x) \right\rangle , \\
&& \lim_{N\rightarrow \infty}
\left\langle q(y) \, \bar A^a_{\mu {\rm DW}}(x) \, \bar q(z) 
\right\rangle_{\rm c}    \nonumber\\
&& \qquad \quad 
=Z \left\langle \psi(y) \, \widetilde A^a_{\mu {\rm GW}}(x) \, 
\bar \psi(z) \right\rangle_{\rm c} .  
\end{eqnarray}
On the other hand, the term which breaks chiral symmetry in the axial
Ward-Takahashi identity is given in terms of the heavy flavors:
\begin{equation}
\label{eq:axial-WT-identity-DW}
 \partial_\mu^\ast \left\langle \bar A_{\mu {\rm DW}}(x) \right\rangle 
= 
\frac{2}{a_5}
\left\langle \bar \psi^\prime_{\frac{N}{2}}(x) \, \gamma_5 \, 
\psi^\prime_{\frac{N}{2}}(x) \right\rangle + \cdots .
\end{equation}
The r.h.s. times $a^4$ can be evaluated as 
\begin{eqnarray}
({\rm r.h.s.})&=& - a {\rm tr} \gamma_5 R D_N (x,x)  \\
&& 
- {\rm tr}\left(
 D_N^{-1} \gamma_5 \frac{1}{a} 
                   \frac{1}{\cosh^2 \frac{N}{2} a_5 \widetilde H}
\right)(x,x).  
\nonumber
\end{eqnarray}
Then we can see that the chiral symmetry breaking reduces to 
the chiral anomaly of the Ginsparg-Wilson fermion 
in the limit $N\rightarrow \infty$, 
\begin{equation}
 - a {\rm tr} \gamma_5 R \widetilde D(x,x).
\end{equation}

\section{Conclusion}

In conclusion, 
the light fermion field 
introduced by Furman and Shamir 
has a rather direct and simple relation to the Ginsparg-Wilson 
fermion field described by the (truncated) overlap Dirac operator.
The chiral properties of the low energy observables of the 
domain-wall fermion can be interpreted in terms of the exact 
chiral symmetry based on the Ginsparg-Wilson relation.

This direct relation can be extended to the case of chiral fermions: 
coupled to an interpolating five-dimensional gauge field, 
the complex phase of the partition function of the domain-wall fermion
can be regarded as a lattice implementation of the eta-invariant 
and it has a direct relation to 
the effective action of the chiral Ginsparg-Wilson fermion, 
which is defined through the chirality $\hat \gamma_5=\gamma_5(1-aRD)$. 
See \cite{aoyama-kikukawa} for detail.

\end{document}